\documentclass[a4paper]{spie325}  
\usepackage{graphicx}
\usepackage{amsmath}
\usepackage{subfigure}
\usepackage{url}
\title{The Fringe Detection Laser Metrology for the GRAVITY Interferometer at the VLTI}

\author{H.~Bartko\supit{a}, S.~Gillessen\supit{a}, S.~Rabien\supit{a}, M.~Thiel\supit{a}, A.~Gr\"ater\supit{a}, M.~Haug\supit{a}, S.~Kellner\supit{a}, F.~Eisenhauer\supit{a},
S.~Lacour\supit{b}, C.~Straubmeier\supit{c}, J.-P.~Berger\supit{d,e}, L.~Jocou\supit{d}, W.~Chibani\supit{a}, S.~L\"ust\supit{a}, D.~Moch\supit{a}, O. Pfuhl\supit{a}, W.~Fabian\supit{a}, C.~Araujo-Hauck\supit{c}, K.~Perraut\supit{d}, W.~Brandner\supit{f}, G.~Perrin\supit{b}, A.~Amorim\supit{g}
\skiplinehalf
\supit{a} Max-Planck-Institute for Extraterrestrial Physics, Garching, Germany; \\
\supit{b} LESIA, UMR 8109, Observatoire de Paris, CNRS, UPMC, Universit\'e Paris Diderot, 92190 Meudon, France; Groupement d'Int\'er\^et Scientifique PHASE (Partenariat Haute r\'esolution Angulaire Sol Espace) between ONERA, Observatoire de Paris, CNRS and Universit\'e Paris Diderot;\\
\supit{c} I. Physikalisches Institut, Universit\"at zu K\"oln, 50937 K\"oln, Germany; \\
\supit{d} Laboratoire d'Astrophysique, Observatoire de Grenoble, 38041 Grenoble C\'edex 9, Frances; \\
\supit{e} ESO, Garching, Germany; \\
\supit{f} Max-Planck-Institut f\"ur Astronomie, 69117 Heidelberg, Germany; \\
\supit{g} SIM, Faculdade de Ciˆencias da Universidade de Lisboa,
Portugal. }

\authorinfo{Send correspondence to H. Bartko (E-mail: hbartko@mpe.mpg.de)}

\begin{document}
  \maketitle

\begin{abstract}
Interferometric measurements of optical path length differences of stars
over large baselines can deliver extremely accurate astrometric
data. The interferometer GRAVITY will simultaneously measure two
objects in the field of view of the Very Large Telescope
Interferometer (VLTI) of the European Southern Observatory (ESO) and
determine their angular separation to a precision of 10 $\mu$as in
only 5 minutes. To perform the astrometric measurement with such a
high accuracy, the differential path length through the VLTI and the
instrument has to be measured (and tracked since Earth's rotation
will permanently change it) by a laser metrology to an even higher
level of accuracy (corresponding to 1 nm in 3 minutes). Usually,
heterodyne differential path techniques are used for nanometer
precision measurements, but with these methods it is difficult to
track the full beam size and to follow the light path up to the
primary mirror of the telescope. Here, we present the preliminary
design of a differential path metrology system, developed within the
GRAVITY project. It measures the instrumental differential path over
the full pupil size and up to the entrance pupil location. The
differential phase is measured by detecting the laser fringe pattern
both on the telescopes' secondary mirrors as well as after
reflection at the primary mirror. Based on our proposed design we
evaluate the phase measurement accuracy based on a full budget of
possible statistical and systematic errors. We show that this
metrology design fulfills the high precision requirement of GRAVITY.
\end{abstract}



\section{Introduction}
\label{sec:intro}  

\begin{figure}[ht!]
\begin{center}
\includegraphics[totalheight=18cm]{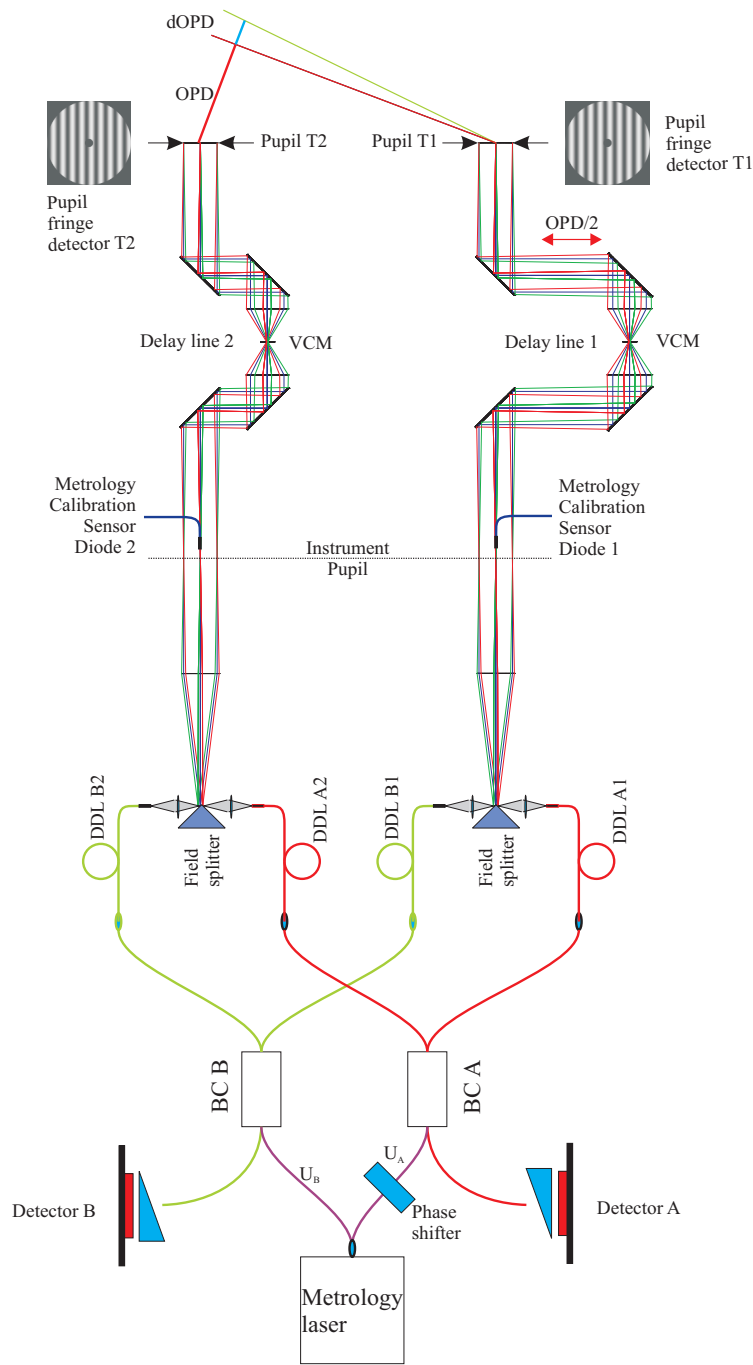}
\caption{Principle of the GRAVITY metrology system shown for two
telescopes. A laser beam travels the same path backwards through all
the beam combination and the VLTI optical train. Detection of the
resulting fringe pattern takes place by imaging the VLT secondary
mirror plus dedicated scattering screens on the telescope spider
with an off-axis infrared camera.} \label{fig:Metrology_overview}
\end{center}
\end{figure}

Interferometric measurements of path length differences of stars
over large baselines can deliver extremely accurate astrometric
data, which is one of the motivations to build an astronomical
interferometer. The interferometer GRAVITY will be capable of
detecting two objects simultaneously in the field of view of the
VLTI. Both objects are separated at the VLTI interferometric
laboratory within the instrument and detected in separate beam
combiners. To perform the astrometric measurement with high
accuracy, the differential path length through the VLTI and the
instrument has to be known (and tracked since Earth's rotation will
permanently change it) at an even higher level of accuracy. This is
the scope of a dedicated system, commonly called metrology. The
error budget of the whole instrument leaves an RMS dOPD
(differential optical path difference) error of 1~nm within a total
of 3~min integration time for the metrology (see Gillessen et
al.\cite{Gillessen2010} 2010). The difference in path length occurs
primarily within the instrument after the field separation. But a
difference in path length can also occur within the VLTI, 
since the traveled path depends on the exact footprint of the beam through the VLTI optics.
Special points where large systematic differences
can occur are at the field locations, where optical elements are
present. The variable curvature mirror (VCM) with $\sim$150~nm surface
variation due to adaption of the surface shape is among the
candidates to introduce large field dependent path length changes in
a systematic way. Thus an active measurement of the path length
within the instrument and the VLTI is required. The PRIMA facility
uses a heterodyne difference frequency technique with lasers
launched at the center of the pupil at the beam combiner and being
back reflected from a retro reflector behind the MACAO dichroic
mirror (Leveque et al.\cite{Leveque2003} 2003). The quantity
measured is the optical path length difference for each object
between the two telescopes from the beam combiner to the dichroic
mirror. On the one hand, the difference frequency technique is known
to be extremely accurate and laser intensity independent. On the
other hand, the center of pupil measurement does not trace the path
of the science light, since soon after leaving the pupil
plane the two beams begin to travel along different paths and
measuring the center of the beams will not represent the complete
truth. In particular, in a focal plane the small laser beam will
form a larger PSF than the science light does. The preliminary
design of the GRAVITY metrology follows an implementation which
(Rabien et al.\cite{Rabien2008} 2008):
\begin{itemize}
 \item {maps the full pupil with the metrology beams}
 \item {covers the total length of the instrument and the VLTI up to the entrance pupil of
the system, the secondary mirror of the VLT telescopes}
 \item {allows a measurement for all beams in parallel}
 \item{requires only minor retro fittings of the VLTI or VLTs.}
\end{itemize}
The basic idea is to split a laser beam into two equally bright
beams with a fixed phase relation and inject them backwards into the
two beam combiners. From both beam combiners four beams will then
travel the exact same path backwards through the instrument and VLTI
as the science light is traveling forward. At the VLT secondary, the
entrance pupil of the system, the laser beams from the two beam
combiners will interfere and form a fringe pattern. This fringe
pattern will be detected with an off-axis infrared camera in the
scattered light from the UTs secondary mirror plus dedicated
scattering screens attached to the telescope spider holding the
secondary mirror. The interferogram can finally be used to determine
the differential path up to this point. This concept is illustrated
in Figure \ref{fig:Metrology_overview}.

\section{Physical Requirements on GRAVITY Metrology}
\label{sec:intro}  

\subsection{Accuracy of Phase Measurements}\label{sec:accuracy_requirements}

The required accuracy for the metrology measurement is driven by
three factors:
\begin{itemize}
\item{The measurements of the stellar phase differences should be the main
contributors to the astrometric error budget; the metrology error should only contribute marginally: 1~nm dOPD error in 3 minutes}
\item{The metrology signal shall allow the differential delay lines (DDL) loop to be stably closed: maximum phase error per ABCD 0.13~rad, see section \ref{sec:ddl_loop_stability}.}
\item{The metrology needs to be fast enough to drive the phase at
a reasonable speed during the acquisition procedure.}
\end{itemize}
%

\subsection{Metrology Fringe Detection in M1 Space}

As shown by Lacour et al.\cite{Lacour2010} (2010), the narrow angle
baseline for the astrometric mode of GRAVITY is defined by the
vector between the two points (projected to M1 space if not defined
therein) on each telescope where the metrology phase measurement is
made. This requires a metrology fringe detection after reflection on
M1 at a fixed position with respect to M1.

For both UTs and ATs we will sample the metrology interference
pattern after reflection on M1 with dedicated scattering screens /
photo diodes, which have fixed positions with respect to M1 on the
M2/M3 spider arms. We use their signals to compute the metrology
phase in real-time. All individual phase measurements will be
weighted and averaged to obtain a single phase measurement per
telescope.

In our current baseline design, as presented in this document, we plan to record full images of the
fringe pattern on M2 with IR cameras in scattered light for the UTs. This data will enable us to correct off-line for possible higher order aberrations (Rabien
et al.\cite{Rabien2008} 2008). In August 2010 we will conduct a Metrology test at the ESO Paranal observatory to measure the amount of quasi-static differential abberations, and their stability. In case of sufficiently small and stable abberations, it may be sufficient to only sample the metrology fringe pattern with diodes also in the case of the UTs.

\section{Metrology System Design}

\subsection{Basis Description of Working
Principle}\label{sec:metrology_basics}

Figure 1 shows a schematic drawing of the optical path within the
VLTI and the GRAVITY instrument. The light from two astronomical
objects enters the system at the telescope pupils, e.g. the VLT
secondary mirrors. The brighter of the two objects (A) is marked in
red, the other one (B) in green. The optical path length difference
to the beam combination has to be compensated for with the VLTI main
delay lines. Within the instrument the light from the two objects is
separated at the field splitter and sent through differential delay
lines (DDLs) to the beam combination (BC A and BC B), after which
the light is fed into the spectrometers. A laser source operating at
1908 nm near the atmospheric K band is split into two beams and sent backwards into
the beam combiners. Starting at the beam combiner, the laser light
travels exactly the same path backwards through the VLTI, as the
light from objects A and B travels forward. At each pupil the laser
wavefronts overlap, tilted by an angle with respect to each other
that corresponds to the angular separation of the astronomical
objects A and B, and create interference fringes. The maximum number
of fringes for a 2'' UT field of view (FOV) and a metrology wavelength of 1908 nm is
40.5 fringes. These fringes can be detected in scattered light from
the secondary mirror, as well as from dedicated scattering screens
mounted on the telescope spider, with an off-axis infrared camera,
or directly using infrared sensitive photodiodes. The following
analysis of the optical path length differences along the science
and the metrology beams shows that the proposed setup is able to
deliver the desired differential dOPD from the measurements on sky
and from the fringe pattern detected on the telescopes. The
individual optical path lengths through the system are denoted as:

\begin{center}
\begin{table}[!h]
\begin{center}
\small
\begin{tabular}{l l}
$L_{A1}$ & red path from the telescope T1 to the DDLA1 \\
$L_{A2}$ & red path from the telescope T2 to the DDLA2 \\
$L_{B1}$ & green path from the telescope T1 to the DDLB1 \\
$L_{B2}$ & green path from the telescope T2 to the DDLB2 \\
$DL_{xx}$ & path length from the differential delay line entrance up
to
the beam combination \\
$U_A$ , $U_B$ & path lengths from the metrology laser splitter to
the beam combiners.
\end{tabular}
\end{center}
\end{table}
\end{center}

The following four quantities $\Psi_A$, $\Psi_B$, $\Phi_A$ and
$\Phi_B$ are measured: $\Psi_A$ is the phase difference of the light
of star A with an effective wavelength $\lambda_A$ measured at
detector A (for the definition and measurement of effective
wavelength see Choquet et al.\cite{Choquet2010} 2010):
\begin{equation}
\Psi_A = \left\{ (L_{A1}+DL_{A1}) - (L_{A2}+DL_{A2}+OPD) \right\}
\frac{2\pi}{\lambda_A} \ .
\end{equation}
$\Psi_B$ is the corresponding phase difference of the light of star
B measured at detector B. Since the phases of the stars' light will
be measured for multiple wavelengths, an absolute length difference
can be detected, e.g. the center of the white light fringe can be
determined at the point where $\Psi_A=0$ for all $\lambda_i$ (see
Choquet et al.\cite{Choquet2010} 2010). In contrast to that, the
phase difference for the single metrology laser wavelength is only
unambiguous within the interval $-\pi < \Phi < \pi$. At the laser
fringe detector at telescope number 1 the phase difference of the
laser beams for the laser wavelength $\lambda_L$ is:
\begin{equation}
\Phi_1 = \left\{ (L_{A1} + DL_{A1} + U_A) -
(L_{B1}+DL_{B1}+U_B)\right\} \frac{2\pi}{\lambda_L} + n_1 2 \pi =
\left\{ (L_{A1}+DL_{A1})-(L_{B1}+DL_{B1})\right\}
\frac{2\pi}{\lambda_L} + \Delta_1 \ ,
\end{equation}
similar at the laser fringe detector number 2:
\begin{equation}
\Phi_2 = \left\{ (L_{A2} + DL_{A2} + U_A) -
(L_{B2}+DL_{B2}+U_B)\right\} \frac{2\pi}{\lambda_L} + n_2 2 \pi =
\left\{ (L_{A2}+DL_{A2})-(L_{B2}+DL_{B2})\right\}
\frac{2\pi}{\lambda_L} + \Delta_2 \ .
\end{equation}
The constants $\Delta_1$ and $\Delta_2$ can be determined with the
metrology zero point calibration procedure, see section
\ref{sec:zero_point}. In the astrometric mode of GRAVITY the
quantity that has to be measured is dOPD on sky:
\begin{equation}
\mathrm{dOPD} = \Psi_A\frac{\lambda_A}{2\pi} -
\Psi_B\frac{\lambda_B}{2\pi}+(\Phi_2-\Phi_1)\frac{\lambda_L}{2\pi} -
(\Delta_2-\Delta_1)\frac{\lambda_L}{2\pi} \  . \label{eq:Deltas}
\end{equation}

This equation contains only the measurable quantities and the
calibration phases. It shows that the quantity dOPD can be retrieved
from the beam combiners and the metrology signals. Each of the
values above contributes to the final error budget.

\subsection{Phase Shifting Interferometer}\label{sec:phase_shifting_interferometer}

The laser wavefronts that are launched from the fiber exits will
overlap in the pupil plane of the telescopes. From the resulting
intensity distribution the global phase offset has to be extracted.
The intensity distribution is detected with a camera which images
the scattered light from the secondary mirror and dedicated
scattering screens installed on the telescope spiders. The phase
extraction from these images has to solve for the following properties:
\begin{itemize}
\item{Most likely the intensity distribution, even if being
emitted perfectly Gaussian from the fibers, will not be smooth and
Gaussian any more after having passed all the VLTI beam train.}
\item{The scattering process from the solid surface and dust particles
will result in a speckle pattern on the camera, due to the coherence
of the light and the surface inhomogeneities.}
\item{The intensity of the scattered light at the secondary mirrors will
be low, since it is a high quality mirror.}
\item{There will be background radiation, since the ambient temperature is
280K. }
\item{The intensity pattern will rotate with the field
rotation on sky.}
\item{ During the calibration process the number
of fringes in the pupil plane will vary between 0 and 40 for a
maximum separation of the objects of 2'' (UTs).}
\item{The phase computation must be sufficiently fast to steer the differential
delay lines (see section \ref{sec:zero_point})}.
\end{itemize}

To meet these requirements the proposed method for the phase
extraction follows a routine based on phase shifting interferometry.
Widely used for the measurement of optical surfaces, phase shifting
interferometry (PSI) allows a phase difference to be measured with
high accuracy. Standard laboratory interferometers make use of the
comparison between a reference phase and a phase under test, e.g. in
a Twyman-Green setup. The reference beam is phase shifted in several
steps, and an interferogram is recorded at each step. Due to the
coherence of the laser wavefronts, the intensity at each position
$(x,y)$ follows the equation of a two-beam interference:
\begin{equation}
I(x,y,\alpha) = I'(x,y)+I''(x,y) \cos(\Phi(x,y)+\alpha) \ .
\end{equation}
Here, $\alpha$ is the global additional phase difference, which is
deterministically added to the launched beam. There are several
standard algorithms to calculate the local phase from a set of phase
shifted interferograms, which can be found in textbooks (e.g.
Malacara\cite{Malacara1992} 1992). With the four-step algorithm four
interferograms are recorded with phaseshifts of
$\alpha_i={0,\pi/2,\pi,3\pi/2}$ respectively. The solution to the
four equations contains only the measured intensities:
\begin{equation}
\tan(\Phi(x,y)) = \frac{I_4 - I_2}{I_1 - I_3} \  .
\end{equation}
Comparable solutions can be found for three or more steps, e.g. the
Hariharan or Carr\'e algorithm. Since we are only interested in a
global phase difference from the calibration point to the
observation location in the field, an absolute phase measurement is
not needed, and the observed $2\pi$ jumps can be recorded in a
counter. The ABCD algorithm allows a relatively simple and straight
forward calibration of the phase shifter to the values even for a
non-linear phase shift response of the phase shifter to the control
voltage, see section \ref{sec:phase_shifter_calibration}.

\section{Hardware}

\begin{figure}[ht!]
\begin{center}
\includegraphics[width=\textwidth]{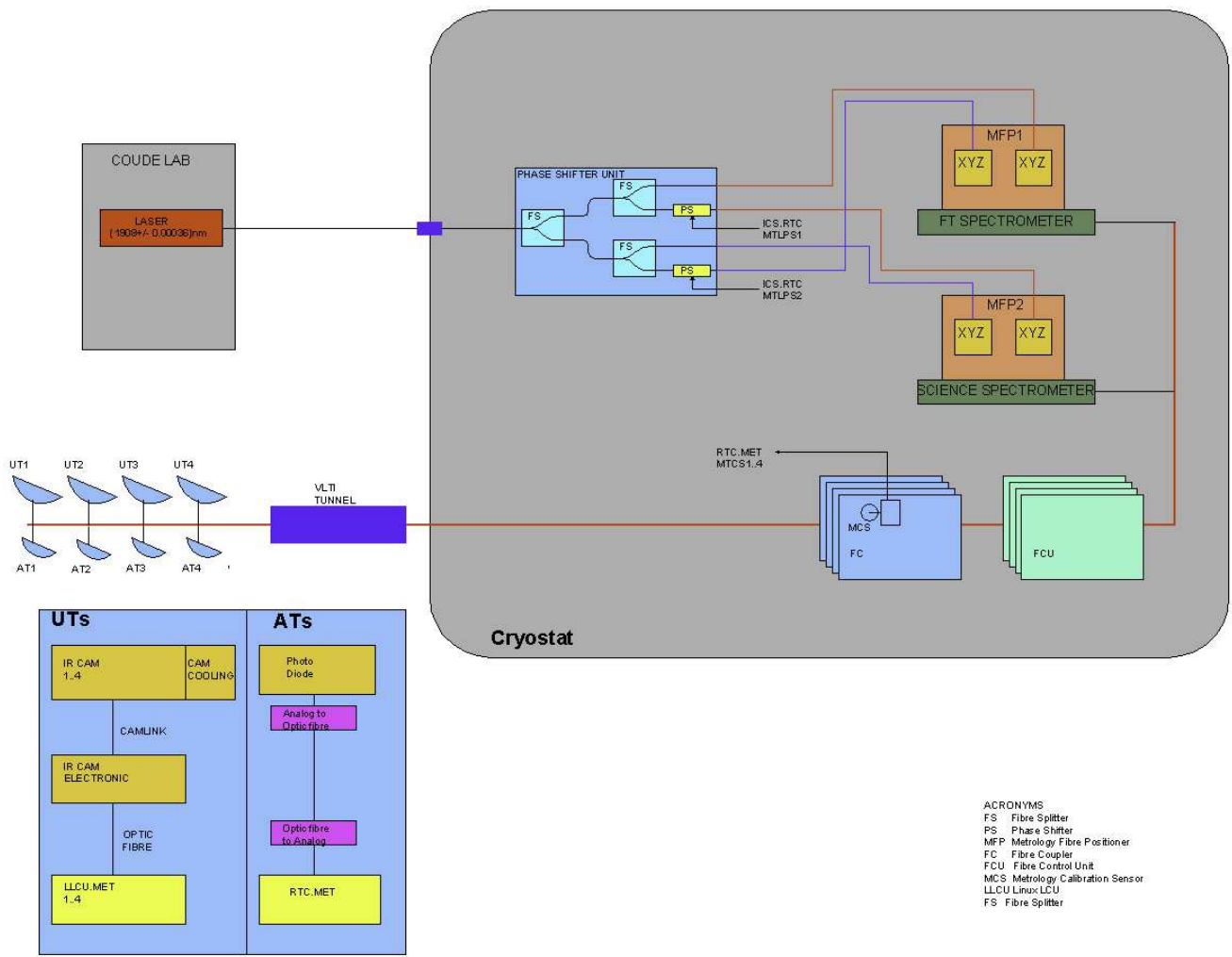}
\caption{Sketch of the metrology hardware. The metrology laser
system is located in the electronics racks of the Combined Coud\'e
laboratory. The metrology laser light is sent via a single mode
polarization maintaining fiber to the GRAVITY cryostat. In the
cryostat a splitting and phase shifting unit is located. The laser
is first split in two steps into four fibers containing equal
amounts of light. Two of the fibers pass an electro-optical phase
shifter before they are directed to the metrology fiber positioners.
The two other fibers are directly routed to the fiber positioner.
The fibers end at the interface between the metrology fiber
positioner and the cryostat. The metrology light enters the laser
collimation optics of the spectrometer as a free beam. It is
directed towards the IO beam combiner and overlayed onto the stellar
light with a dichroic beam splitter. Finally, the metrology laser
light is fed into two of the 24 exits of the integrated optics. In
the IO the laser light is split again and distributed to the four
entrance fibers of the IO beam combiners. The interference pattern
in the pupil of the Fiber Coupler is sampled by the metrology
calibration sensor diode and allows an accurate calibration and
monitoring of the active components like the metrology laser, phase
shifter and fiber controllers. A characteristic interference pattern
of the metrology laser light will be visible at the entrance pupil
of the telescopes, M2. In case of the UTs the full M2 pupil will be
imaged in scattered light, plus dedicated scattering screens on the
M2 spider. In case of the ATs the pupil will be sampled with
photodiodes mounted on the M3 spider.}
\label{fig:Metrology_hardware}
\end{center}
\end{figure}

In this section we will describe the preliminary design and layout
of the hardware of the GRAVITY metrology. The analysis of the
performance of the full GRAVITY metrology system based on the
hardware requirements will be presented in section
\ref{sec:system_performance}.

\subsection{Hardware Overview}

Figure 2 shows a sketch of the metrology hardware: The hardware of
the metrology system located in or near the VLTI laboratory includes
the metrology laser source, the beam transmission fiber, the
splitting and phase shifting unit and the metrology fiber positioner
for the coupling adjustment of the IR laser light into the IO via
the spectrometer optics. The laser itself with the frequency control
unit will be located in the GRAVITY electronics racks in the
Combined Coud\'e lab. The laser foreseen will be a commercial high
power single frequency laser. The delivery of the laser light to the
GRAVITY beam combiner instrument will be done with a single mode
polarization maintaining fiber. Since the phase of the laser in
front of the cryostat is unimportant, the fiber delivery needs no
special protection. The fiber is connected to the inside of the
cryostat by a commercially available single mode and polarization
maintaining fiber feed-through.

In the cryostat the laser will be split into four beams with equal
intensity. Two of the arms are directed to electro-optical phase
shifters. After the phase shifting the four fibers are routed to the
metrology fiber positioners, which align the fibers in the field
before the fiber coupling optics of the spectrometer (Araujo-Hauck
et al.\cite{AraujoHauck2010} 2010, Straubmeier et
al.\cite{Straubmeier2010} 2010). The alignment is motorized in three
axes for each fiber to allow a remote coupling efficiency
optimization, while the cryostat is closed. After the light leaves
the fibers, it is collimated at the laser collimator in the
spectrometer and joins the science light at a dichroic beam splitter
and is injected into the waveguide of the integrated optics (IO,
Jocou et al.\cite{Jocou2010} 2010).

The metrology light passes the integrated optics backwards, in the
opposite direction of the starlight. Inside the IO the metrology
light is split such that all four entrance fibers of the IO contain
equal amounts of metrology light. The metrology light travels trough
the fiber control units containing the differential delay
lines (DDLs) and polarization rotators (see Perrin et
al.\cite{Perrin2010} 2010) to the field splitter of the GRAVITY
fiber coupler (see figure \ref{fig:Metrology_calibration_diode}). In
the field splitter the metrology light from the beam combiners A and
B is Combined and via the VLTI optical train guided to the four
telescopes. As the laser light from the two beam combiners A and B
is coherent, there is a characteristic interference pattern in each
instrument pupil. The number of fringes visible in the instrument
pupil depends on the angular separation of the fibers in the fiber
coupler projected on the sky. The 2" maximum field of view of the
VLTI (UTs) corresponds to 40 interference fringes.

For the GRAVITY metrology two instrument pupils are of special
importance: first, the pupil in the GRAVITY fiber coupler (Pfuhl et
al.\cite{Pfuhl2010} 2010). The interference pattern in this pupil is
sampled by the metrology calibration sensor diode and allows an
accurate calibration and monitoring of the active components like
the metrology laser, phase shifter and fiber controllers. Second,
there is also the characteristic interference pattern of the
metrology laser light visible at the entrance pupil of the
telescopes, M2. In case of the UTs the full M2 pupil will be imaged
in scattered light, plus dedicated scattering screens on the M2
spider. In case of the ATs the pupil will be sampled with
photodiodes mounted on the M3 spider.

%

In the following we will detail the hardware items and specify their
performance requirements.

\subsection{IR Laser System}

The metrology laser system shall produce high power single frequency
laser light coupled inside a single mode polarization maintaining
fiber. The stability of the laser frequency as well as the laser
power is of special concern. The proposed laser system shall fulfill
the following requirements:
\begin{itemize}
\item{power output: $>2$W continuous wave}
\item{single frequency laser with output wavelength: $(1908 \pm 1)$ nm, stabilized to $\pm30$~MHz absolute wavelength accuracy} 
\item{linear polarized}
\item{power output fluctuations: $<0.5\%$~RMS  of the integrated laser power
  over a 33 ms timescale.}
\end{itemize}

\subsection{Passive Fiber Optic
Components}\label{sec:passive_fiber_components}

The polarization maintaining fibers transport the laser light from
the laser source to the metrology fiber positioner units. The fiber
path from the metrology laser splitter to the beam combiner,
$U_{A/B}$ is of special importance. It is a so-called non-common
path, where the laser light does not follow the star light. The
path length difference $\Delta_2-\Delta_1$ (see equation
\ref{eq:Deltas} and section \ref{eq:Deltas}) can be calibrated, see
equation \ref{eq:delta_calibration} and section
\ref{sec:zero_point}, but needs to be stable at a level below 0.5~nm
during the observation (see section
\ref{sec:phase_extraction_budget}). The optical path length of an
optical fiber changes ($\Delta L$) as a function of temperature
change ($\Delta T$):
\begin{equation}
\Delta L = (n \alpha + \mathrm{d}n/\mathrm{d}T) l \Delta T \ ,
\end{equation}
where $\alpha$ denotes the linear thermal expansion coefficient, $n$
the index of refraction and $l$ is the fiber length. Standard fused
silica fibers show an optical path length change of: $\Delta L =
1\mathrm{nm} (\Delta T / 10\mathrm{mK}) (l/1\mathrm{cm})$ (Leviton
\& Frey\cite{Leviton2006} 2006). Air-guiding photonic-bandgap fibers
offer much smaller path length differences (Dangui et
al.\cite{Dangui2005} 2005).

\subsection{Phase Shifter} \label{sec:phase_shifter_hardware}

The two phase shifters provide the phase shifts of
$\alpha_i={0,\pi/2,\pi,3\pi/2}$  necessary for the ABCD phase
shifting interferometer algorithm. For this a very accurate
reproducibility of the phase shifts is of utmost importance. The
phase shifters are commercially available as integrated optics
components employing LiNbO3 crystals. By applying an appropriate
voltage to the birefringent crystal a corresponding change in the
extraordinary refractive index will occur. If the optical input is
both linearly polarized and aligned with the extraordinary axis of
the modulator crystal, the output will undergo a pure phase shift
with no change in the state of polarization. The phase shifters
shall fulfill the following requirements:
\begin{itemize}
\item{transmission $>70\%$, full-wave amplitude modulation $<0.5\%$}
\item{phase shifting repeatability $<\lambda/5000$}
\item{operation laser power: 0.5 W, damage threshold $>1$~W}
\item{min.~13 effective bits, absolute voltage repeatability $<1/5000$.}
\end{itemize}

\subsection{Metrology Calibration Sensor
Diode}\label{sec:calibration_diode_hardware}

The metrology Calibration Sensor Diode samples the light flux in the
center of the 18 mm diameter pupil in the Fiber Coupler subsystem
(see figure \ref{fig:Metrology_calibration_diode}).  This allows to
(see also section \ref{sec:calibration_diode}):
\begin{itemize}
\item{adjust/monitor the metrology fiber positioner}
\item{monitor the laser power}
\item{in-situ calibrate the phase shifter}
\item{calibrate the DDL, monitor the DDL position}.
\end{itemize}
Requirements:
\begin{itemize}
\item{pick-up fiber with core diameter: 100...110 $\mu$m, total transmission T$>90\%$}
\item{combined diode and amplifier responsivity: $8\cdot10^{-9}$ W
shall be converted to 10 V, bandwidth (3dB) 1000~Hz, input current
equivalent noise: $< 3\cdot10^{-12}$ A / Hz$^{1/2}$, maximum
non-linearity for 10 V output of $<0.1\%$}
\item{ADC sampling rate $\geq1$kS/s, min. 12 effective bits, absolute precision
 $< 1/1000$.}
\end{itemize}

\subsection{Metrology Light Detection at UTs}

The UT IR receivers consist of an IR camera and optics each. Here,
we describe the design parameters of both components. The detection
will be two-fold: The first order signal will come from scatterers
mounted onto the M2-spider (section \ref{sec:scattering_screens}),
the higher order corrections can be derived from the image of M2.
Both signals will be measured with the same cameras.

\subsubsection{IR Cameras}

From the physical requirements, we derive the following requirements
for the IR cameras:
\begin{itemize}
\item{The image of M2 needs to be mapped onto at least $160 \times
160$ pixels (4 pixels per fringe, 40 fringes).}
\item{It shall be possible to observe the inner parts of
the spider arms as well.}
\item{The read noise should be below 250 electrons.}
\item{A quantum efficiency of at least $30\%$ at $1.9 \mu$m.}
\item{The frame rate shall be adjustable from 10 Hz to 200 Hz.}
\item{The dynamic range must be sufficient to allow for exposures
of 40ms at least at the given thermal background.}
\end{itemize}

\subsubsection{Scattering Screens on Telescope Spider}\label{sec:scattering_screens}

The IR cameras will also be used to image the (inner parts of the)
spider arms on which M2 is mounted. Optically, these arms are in the
M1 space. The metrology light gets reflected via M2 and M1 into the
direction of the two celestial objects that are being observed with
GRAVITY. This light also shines on the spider arms from below.
Hence, having a fix reference point on the spider arms at which the
ABCD algorithm can be observed with the cameras is a suitable
realization of the requirement that the metrology signal shall be
measured in M1 space (Lacour et al.\cite{Lacour2010} 2010). The
fixed reference points can be optimized to be good scatterers to
ease detection of the signal. The scattering screens on the
telescope spider shall fulfill the following requirements:
\begin{itemize}
\item{Size: smaller than 1/4 of a fringe. Given the diameter of M1 (8m),
the maximum number of fringes (40), this means the scatterers shall
not exceed 5 cm in diameter.}
\item{The scatterers shall be completely passive devices with a high
scattering strength (the albedo shall be larger than $80\%$).}
\item{The scatterers shall be suitable for detection with an
external metrology, i.e. they should carry clearly visible optical
targets.}
\end{itemize}
Our current baseline design consists of 4 little metal cubes with
dimensions of $20 \times 20$~mm. The surface pointing downwards has
a matte white color; the four surfaces pointing to the sides contain
optical targets. The scatterers will be screwed onto the spider arms
in a position that is
\begin{itemize}
\item{close enough to M2, such that it is inside the field of view
of the cameras}
\item{not too close to M2, such that no excess
metrology light from the beam going from M3 to M2 and shining beyond
M2 perturbs the signal of the metrology light traveling via M1.}
\item{not too close to M2, such that none of the 4 scatterers will
be shadowed by the M2 body, i.e. all are visible from the IR
cameras.}
\end{itemize}

\subsection{Metrology Light Detection at ATs}

The metrology fringes at the ATs will be sampled by four IR
sensitive photo diodes mounted on the four spider arms which hold
the M3 mirror. In order to increase the effective area of the diodes
without increasing their noise, the metrology light is focused by a
lens of 10 mm diameter onto the diode.  A pre-amplifier will be
placed right next to the diode and the amplified and low-pass
filtered analog signal will be transported to the Nasmyth
electronics box. There the four analog signals of the four diodes
will be sent via a digital fiber link using the pre-installed
single mode fibers to all AT stations to the central Combined Coud\'e
lab and converted back to an analog signal. In the Combined Coud\'e
lab the 16 analog signals from the four diodes at the four ATs will
be directly digitized by the metrology real-time computer (see section
\ref{sec:computing}).

The metrology fringes detection and read-out on the ATs consists of
the following components:
\begin{itemize}
\item{4 photodiodes on M3 spider + pre-amplifier (max 1 W power
consumption) per AT}
\item{analog optical link between AT stations and Combined Coud\'e lab}
\item{ADC.}
\end{itemize}

The IR sensitive diodes, amplifiers and the ADCs shall fulfill the
same requirements as the respective components of the metrology
calibration sensor diodes, see section
\ref{sec:calibration_diode_hardware}.

\section{Computing / Slow Control} \label{sec:computing}

\begin{figure}[ht!]
\begin{center}
\includegraphics[width=8.7cm]{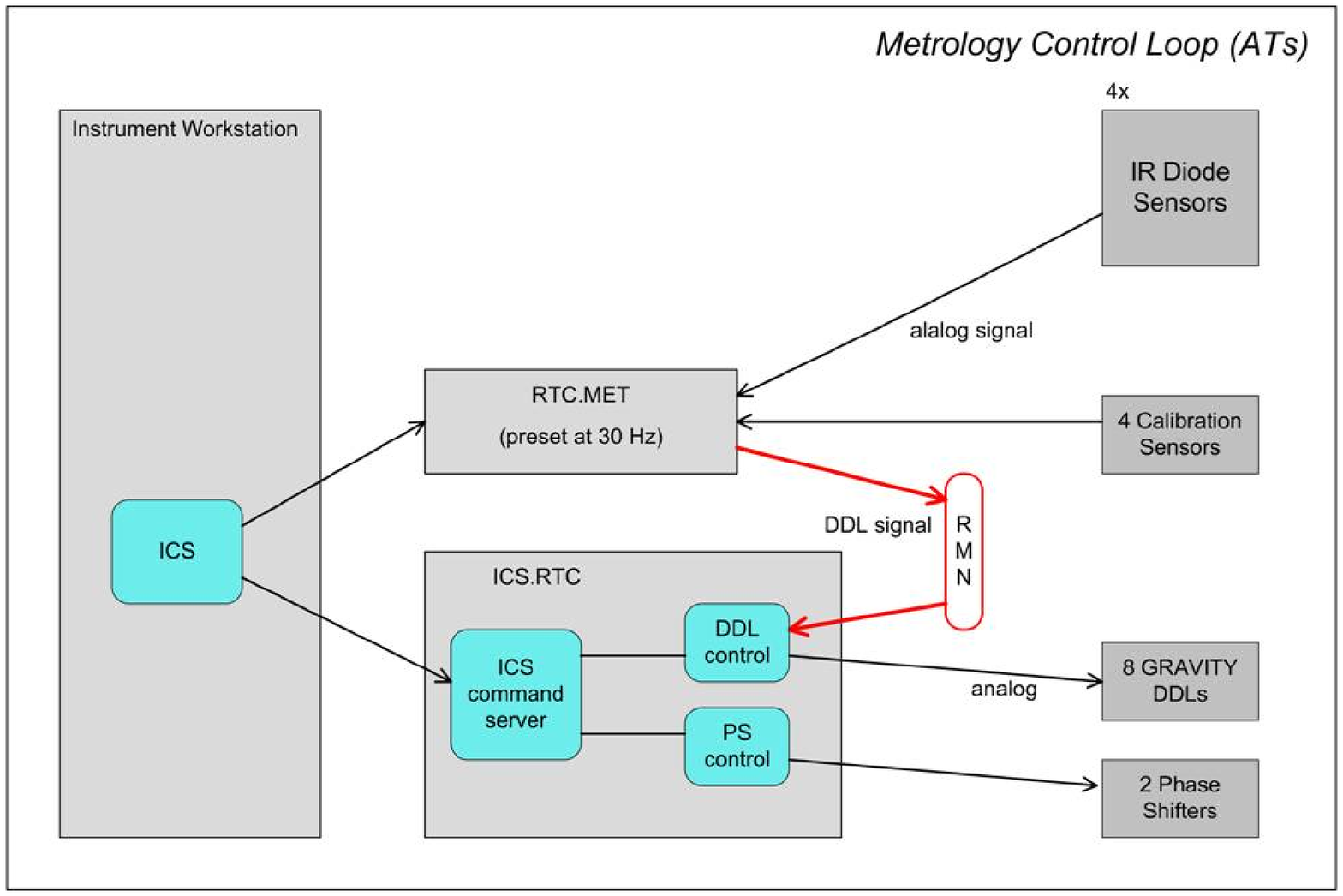}
\includegraphics[width=8.3cm]{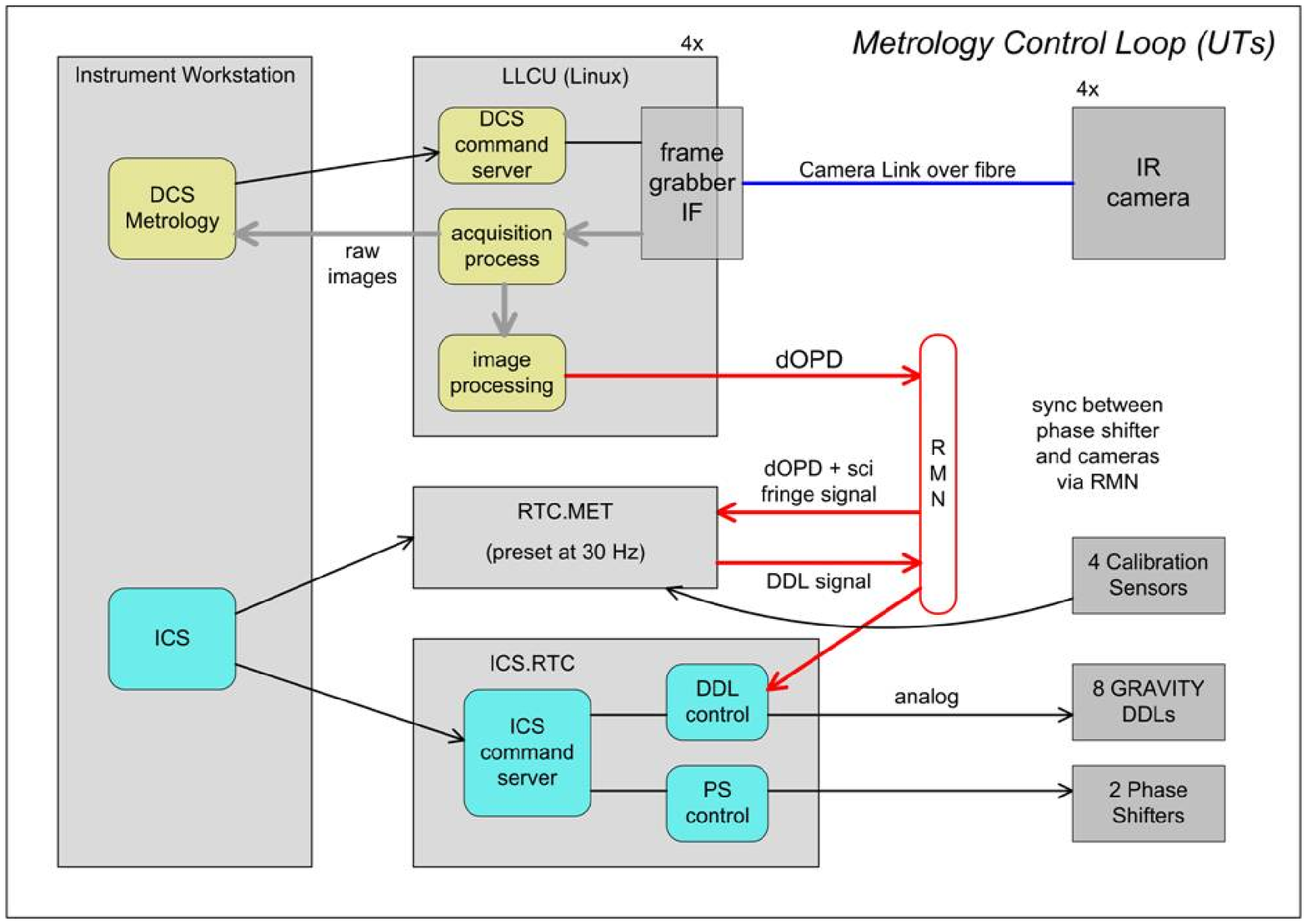}
\caption{Left: Metrology computing block diagram for the case of
ATs. The RTC.MET and ICS.RTC computers are located in the Combined
Coud\'e lab and the IR diodes are located on the M3 spiders of the
ATs. The calibration sensors, DDLs and phase shifters will be
located inside the GRAVITY cryostat in the VLTI lab. Right:
Metrology computing block diagram for the case of UTs. The LLCU,
RTC.MET and ICS.RTC computers are located in the Combined Coud\'e
lab and the IR cameras are located on the UTs. The calibration
sensors, DDLs and phase shifters will be located inside the GRAVITY
cryostat in the VLTI lab.} \label{fig:Metrology_computing}
\end{center}
\end{figure}

The metrology system employs the phase shifting interferometer ABCD
algorithm (see section \ref{sec:phase_shifting_interferometer}) to measure the path difference between
one telescope and the two beam combiners.  This requires an accurate
timing between the phase shifter, the metrology calibration sensor
diode and the metrology fringe detectors at the UTs and ATs. The 30
Hz phase shifting is steered by the metrology real-time local control unit (RTC.MET LCU). This computer is
synchronized via the ESO reflective memory network (RMN, see Wallander, A.\cite{VLTI_computing_ICD} 2006) with the instrument control system real-time control unit (ICS.RTC LCU), which steers the eight DDLs and the two metrology phase
shifters. In the final design phase, we will evaluate whether the
DDLs and phase shifters may also be steered by the RTC.MET LCU
reducing the complexity of the system. The analog signals of the
metrology calibration sensors and the IR sensitive diodes of the ATs
are recorded directly in the RTC.MET LCU. The IR cameras of the UTs
are controlled by a dedicated real time Linux based local control unit (LLCU), which is
connected to the RTC.MET via the RMN. Figure
\ref{fig:Metrology_computing} (left) shows the layout of the
Metrology Control Loop for the case of ATs. Figure
\ref{fig:Metrology_computing} (right) shows the layout of the
Metrology Control Loop for the case of UTs. The raw images of the
metrology cameras are sent directly from the LLCU (including
appropriate time stamp) to the instrument workstation for data
storage.

\subsection{Data Inputs / Outputs}

Table \ref{tab:data_inputs} shows the science data inputs, their
rate and number of bits. All of these data have to be stored
including an accurate time stamp for offline analysis. The data rate
is dominated by the pictures of the IR cameras of the UTs. There are
only two real-time data outputs to steer the phase shifters, 14 bits
each at a rate of 30Hz.

\begin{center}
\begin{table}[!ht]
\begin{center}
\small
\begin{tabular}{c c c}
\hline \hline
Name  &  rate / Hz & $\#$ bits  \\
\hline
laser power & 1  & 12 \\
laser wavelength & 1  & 12 \\
laser status & 1  & 8 \\
$4\times$ metrology calibration sensor (12 bits each) & 1000 & 48 \\
$4\times$ UT IR cameras ($256\times 256$ pixels, 14 bit) & 30 &
4E+06\\
$4\times$4 AT IR diodes (12 bits each) & 1000 & 3072 \\
\hline
\end{tabular}
\caption{Science data inputs, their rate and number of bits.}
\label{tab:data_inputs}
\end{center}
\end{table}
\end{center}

\subsection{Real-Time Computation Tasks}

The following real-time computation tasks are necessary for the
GRAVITY metrology system:
\begin{itemize}
\item{on the LLCU:
\begin{itemize}
\item{IR camera Image Acquisition Process software}
\item{Metrology Phase Analyzer software}
\end{itemize}
}
\item{on the RTC.MET:
\begin{itemize}
\item{DDL Signal Processor software}
\item{metrology calibration sensor read-out software}
\item{AT fringe sampling diodes read-out software}
\end{itemize}
}
\item{on the ICS.RTC:
\begin{itemize}
\item{DDL Controller software}
\item{metrology phase shifter controller software.}
\end{itemize}
}
\end{itemize}

\subsection{Data Storage}

We plan to store the full metrology raw data in order to be able to
do a detailed offline analysis. The amount of data to be stored is
dominated by the image size of the metrology cameras. To reduce the
data stream to be archived, 40 successive metrology camera raw
frames will always be integrated (co-added) into four frames
corresponding to the phase shifter settings A, B, C, and D, which
will then be stored, resulting in a data stream of 480 kB/s.

\section{Calibration Strategy and Performance}

Here we shortly present the calibration strategy and performance of
the GRAVITY metrology and its subcomponents.

\subsection{Metrology Calibration Sensor
Diode}\label{sec:calibration_diode}

\begin{figure}[ht!]
\begin{center}
\includegraphics[totalheight=7cm]{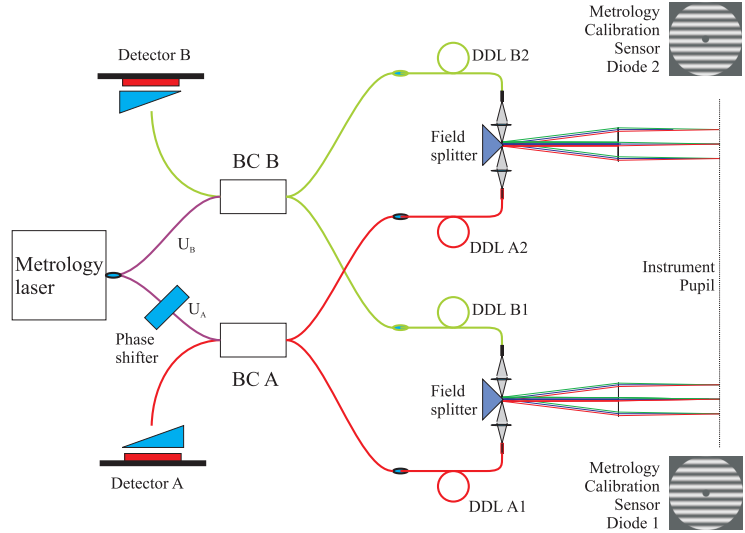}
\caption{Metrology optical path from the IR laser to the Metrology
Calibration Sensor Diode, which samples the interference pattern of
the metrology laser light tracing the path of the two stars A and B
behind the dichroic in the GRAVITY fiber coupler.}
\label{fig:Metrology_calibration_diode}
\end{center}
\end{figure}

The light intensity at the metrology calibration sensor diode 1
mounted in the pupil center is (see figure
\ref{fig:Metrology_calibration_diode}):
\begin{equation}
\Phi_1=\left\{ (DL_{B1}+U_B) - (DL_{A1}+U_A) \right\} 
\frac{2\pi}{\lambda_L} + \alpha(V) \ ,
\label{eq:calibration_diode_phase}
\end{equation}
where $\alpha(V)$ denotes the phase shift of the phase shifter
device according to the control voltage V. Therefore, the metrology
calibration sensor allows to:
\begin{itemize}
\item{monitor the metrology IR laser power}
\item{adjust/monitor the metrology fiber positioner}
\item{in-situ calibrate the phase shifter}
\item{calibrate the DDL, monitor the DDL position.}
\end{itemize}

For a fixed position of the metrology fiber positioners as well as
phase shifter and DDL settings the observed intensity with the
metrology calibration sensor diode is proportional to the power of
the metrology laser.

\subsubsection{Performance of Metrology Calibration Sensor Diode}

According to the metrology transmission budget (see table
\ref{tab:transmission}), we expect an intensity of 18~mW in
the pupil of the fiber coupler. The center of the 18~mm diameter
pupil will be sampled with a fiber of 100~$\mu$m core diameter. The
fiber diameter corresponds to 1/4 of the fringe spacing of max. 40
fringes over the pupil (18mm/40/4 = 112~$\mu$m). The light power in
the fiber is thus (including a transmission of the dichroic of
$1\%$): $I_{\mathrm{fiber}} = 5.6\cdot 10^{-9}$~W. Accounting for
possible losses in the multimode fiber and the coupling between
fiber and metrology calibration sensor photodiode, we expect a light
flux on the diode of $I_{A}+I_B = 5\cdot 10^{-9}$~W. The average
intensity over one ABCD phase shift cycle is then $\langle I_{A}+I_B
\rangle = 2.5\cdot 10^{-9}$~W. According to the specifications (see
section \ref{sec:calibration_diode_hardware}) the error of the
intensity measurement is dominated by the noise of the amplifier of
$< 3\cdot10^{-12}$~A/Hz$^{1/2}$. Together with a response of the
diode of 1~A/W this results in a noise equivalent power per 33~ms
exposure (phase shifter time step) of NEP$ = 1.7\cdot 10^{-11}$W.
According to equation \ref{eq:delta_phi} the phase error
$\Delta\phi$ per 33~ms exposure (calculated from the last 4 ABCD
measurements) is $\Delta\phi =4.3\cdot10^{-3} = 1.5 \ \mathrm{nm}
\frac{2\pi}{\lambda_L}$.

\subsubsection{Phase Shifter Calibration}
\label{sec:phase_shifter_calibration}

Using the GRAVITY metrology as a phase shifting interferometer (see
section 5.2) relies on the phase shifts
$\alpha_i={0,\pi/2,\pi,3\pi/2}$ being exactly known. Errors in
$\alpha_i$ will result in a systematic error of the calculated
phases. The systematic phase offset will be of the same order of
magnitude as the error on $\alpha_i$. Since we want to measure the
phase with the metrology system to 1~nm$\sim \lambda/2000$, the
individual phase shifts have to be known to at least the same level
of accuracy or better. The calibration routine for the phase shifter
developed for this purpose makes use of similar identities as shown
by Rabien et al.\cite{Rabien2006} (2006), see also Rabien et
al.\cite{Rabien2008} (2008).

The method described there requires a linear response of the
metrology calibration sensor diode (see section
\ref{sec:calibration_diode_hardware}). The phase shifter response
function   does not need to be linear. Moreover, the interferometer
needs to be stable during one set of phase shift measurements, e.g.
that the internal dOPD does not drift within a $\lambda/2000$
margin. Since it is possible to measure such datasets fast, this is
not considered to be problematic. For phase shifts of
$\alpha_2-\alpha_1 \sim 90^{\circ}$ the error of the phase shift is
given by:
\begin{equation}
\Delta(\alpha_2-\alpha_1) \sim \frac{\Delta I}{I} \sim
1.8 \ \mathrm{nm}\frac{2\pi}{\lambda_L} \ .
\end{equation}
The ultimate precision of the phase shifter calibration is thus
given by the number of effective bits for the D/A converter
generating the phase shifter voltage. 13 effective bits (see section
\ref{sec:calibration_diode_hardware}) correspond to
$0.23 \ \mathrm{nm}\ 2\pi/\lambda_L$. This calibration precision is
reached after about 60 voltage sweeps. For an experimental test see
Rabien et al.\cite{Rabien2008} (2008).

\subsubsection{DDL Monitoring / Calibration}

The measured phase difference with the metrology calibration sensor
diode is given in equation \ref{eq:calibration_diode_phase}. 
For fixed DDL$_{\mathrm{B1}}$ settings the observed phase shift is
only a function of DDL$_{\mathrm{A1}}$ and the phase shifter setting
(and vice versa). By applying phase shifts
$\alpha_i={0,\pi/2,\pi,3\pi/2}$ one can use the ABCD algorithm to
measure changes in the DDL$_\mathrm{B1}$ length.

\subsection{Metrology Zero-Point Calibration}\label{sec:zero_point}

The strategy for the metrology zero-point calibration is to observe
two nearby bright stars and to swap the two stars from beam combiner
A to B and vice versa. In the first calibration step we observe the
two stars and obtain the following measurements:
$\Psi_{A,\mathrm{cal}},\Psi_{B,\mathrm{cal}},\Phi_{1,\mathrm{cal}},\Phi_{2,\mathrm{cal}}$.
We then swap the two stars from one beam combiner to the other one.
During this step we record the change in the metrology phases,
$\Delta \Phi_{1,\mathrm{cal}}^{\mathrm{swap}}$ and $\Delta
\Phi_{2,\mathrm{cal}}^{\mathrm{swap}}$. Combining these eight
measurements and using equation \ref{eq:Deltas}, we get:
\begin{equation}
\Delta_1-\Delta_2 = \Phi_{1,\mathrm{cal}} -\Phi_{2,\mathrm{cal}} +
\frac{1}{2} \left( \Delta \Phi_{1,\mathrm{cal}}^{\mathrm{swap}} -
\Delta \Phi_{2,\mathrm{cal}}^{\mathrm{swap}} \right) -\frac{1}{2}
\left( \Psi_{A,\mathrm{cal}} + \Psi_{A,\mathrm{cal}}^{\mathrm{swap}}
- \Psi_{B,\mathrm{cal}} -\Psi_{B,\mathrm{cal}}^{\mathrm{swap}}
\right) \frac{\lambda_s}{\lambda_L} \ . \label{eq:delta_calibration}
\end{equation}

\subsection{Metrology Polarization Adjustment}

The whole optical chain from the IR metrology laser to the metrology
fiber positioner is designed with polarization maintaining fibers.
The polarization axis is adjusted to the key of the FC connectors to
within $2^{\circ}$. Also the rotational fiber alignment in the fiber
adapter of the fiber positioner will be done with a maximum
tolerance of $2^{\circ}$. The metrology laser light will be injected
with a high degree of linear polarization along the slow axis of the
IO. In the full GRAVITY setup there will be eight fibered
polarization rotators in front of the IO, see Perrin et
al.\cite{Perrin2010} (2010), one for each of the two stars times
four telescopes. Adjusting the fibered polarization controllers in
front of the IO for maximum fringe contrast on the science targets
needs six degrees of freedom of the eight rotators. A seventh degree
of freedom is sufficient to adjust the relative polarization between
the two stars yielding maximum fringe contrast of the metrology
light at the telescope.

\section{System Performance} \label{sec:system_performance}

In this section we will evaluate the performance expected for the
GRAVITY metrology system based on the above preliminary design. We
will address the following points: transmission budget, DDL control
loop stability, and phase extraction performance.

\subsection{Transmission and Contrast Budget}
\label{sec:transmissioin_budget}

\begin{center}
\begin{table}[!ht]
\begin{center}
\small
\begin{tabular}{c c c c c}
\hline \hline
$\#$ & Name  &  $T[\%]$ 1.9$\mu$m & $\Delta$ T &  Reference \\
&&& BC A/B $T[\%]$ & \\
\hline
1 & vacuum feed-through, splitters, patch-cords & 0.75  & 0.95 / 1.05 &   \\
2 & phase shifter & 0.7 & 1/1 & section \ref{sec:phase_shifter_hardware}\\
3 & 1908nm / K band dichroic & 0.98 & 1/1 & \cite{AraujoHauck2010,Straubmeier2010} \\
4 & laser-IO mode fitting / centering & 0.9 & 0.99/1.01 & \cite{Straubmeier2010}\\
5 & Fresnel losses, scattering, transmission of lenses & 0.9 & 1/1 & \cite{Straubmeier2010}\\
6 & IO 50/50 coupler backward propagation & 0.5 & 1/1 & \cite{Jocou2010} \\
7 & IO 66/33 coupler backward propagation & 0.33 & 1/1 &  \cite{Jocou2010} \\
8 & IO bulk transmission & 0.7 & 0.99/1.01 & \cite{Jocou2010} \\ 
9 & IO-fiber mode fitting / centering & 0.93 & 0.98/1.02 & \cite{Jocou2010}\\
10 & glue transmission & 0.99 & 1/1 & \cite{Jocou2010}\\
11 & fibers and fiber control & 0.9 & 0.98/1.02 & \cite{Perrin2010} \\
12 & end face fiber $\#1$, coated & 0.99 & 1/1& \cite{Perrin2010} \\
13 & fiber coupler & 0.81 & 0.99/1.01 & \cite{Pfuhl2010} \\
14 & VLTI Transmission (UT DF) & 0.28 & 1/1& \cite{VLT_ICD}\\
\hline
   & Total & 0.009 & 0.87/1.13\\
\hline
\end{tabular}
\caption{Transmission budget from the IR metrology laser to the
telescope.} \label{tab:transmission}
\end{center}
\end{table}
\end{center}

For the analysis of the performance of the metrology system the
transmission budget from the IR metrology laser to the telescope is
of special importance, see table \ref{tab:transmission}.
For a 2~W metrology laser, the total intensity (two beams) at the UT
is 4.5 mW. According to Gitton\cite{VLT_ICD} (2010) the VLTI
transmission in the AT DF case is less $(18\%)$ resulting in a laser
power of 2.9 mW at the AT. In case of on-axis observations, half of
the science light is used for the AO (see Gillessen et
al.\cite{Gillessen2010} 2010), such that only half of the metrology
light power can be observed on M2, 2.25mW.

The contrast of the metrology laser fringes at the telescopes is
determined by three effects:
\begin{enumerate}
\item{power differences (0.87/1.13) between the laser light from beam
combiners A and B, see table \ref{tab:transmission}}.
\item{contrast loss due to dOPD change caused by sky rotation during
one metrology exposure (1/(30~Hz)) of max. 78~nm/s or
2.6~nm/exposure, see section \ref{sec:ddl_loop_stability}}
\item{contrast loss due to pupil position
jitter of max. 15 mm, see Pfuhl et~al.\cite{Pfuhl2010} (2010).}
\end{enumerate}
These three effects result in a minimum visibility of the metrology
fringe pattern in M1 space of at least 0.9, which is used in the
simulations of the phase extraction performance in section
\ref{sec:UT_simulations}.

\subsection{DDL Control Loop Stability}
\label{sec:ddl_loop_stability}

\begin{figure}[ht!]
\begin{center}
\includegraphics[totalheight=6cm]{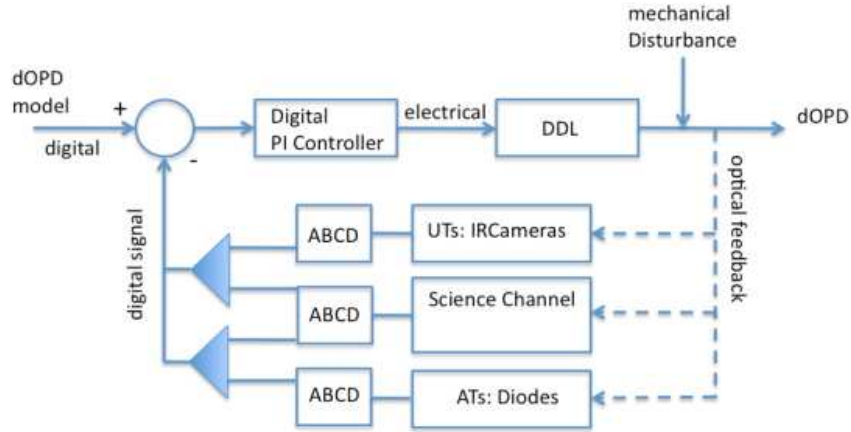}
\caption{Basic design of the control loop for the DDL using the
metrology feedback. Either the UT or AT feedback loop is used. In
both cases, it might happen that the science object is bright enough
to detect the white light fringe online. Then this is the most
immediate feedback and should be used instead.} \label{fig:DDL_loop}
\end{center}
\end{figure}

The metrology signal will be used in two places: In the data
reduction, it is one of the elements delivering the final phase
accuracy. Online, the metrology serves as a measurement device for a
control loop keeping the DDL at the required position. The accuracy
by which the loop has to stay at the desired position (i.e. that the
white-light fringe is matched onto the science channel) is not to be
confused with the astrometric accuracy. For the loop, an accuracy of
$\lambda/10$ is sufficient. This is given by the fact that the
visibility loss due to the jitter in the DDL loop shall not be
limiting. The loop is sketched in Figure \ref{fig:DDL_loop}. For
each observation, the theoretical dOPD value as a function of time
needs to be known. This is a geometrical model, the input of which
is mainly the source coordinates and the wide-angle baseline. While
the coordinates can be safely assumed to be known to sufficient
accuracy, the requirement to be able to track the dOPD translates
into a requirement on how well the baseline is known, see Lacour et
al.\cite{Lacour2010} (2010). Here, the problem is to ensure that the
calculated dOPD track actually is followed by the system, hence dOPD
is the quantity to control. The control loop is a classical feedback
system. The adjustable element is the DDL. As shown by Perrin
et~al.\cite{Perrin2010} (2010), the actual hardware device has a
pronounced hysteresis. This is one of the reasons why dOPD needs to
be controlled and a feed-forward system is not sufficient. The
maximum dOPD occurring is 4~mm, for which the hysteresis reaches
$0.025\% = 1\mu\mathrm{m} \sim \lambda/2$. Even though most of the
time the dOPD values will be much smaller, the design of the system
should under no condition fail due to the hysteresis. The feed-back
of the system is primarily optical. This means that dOPD is measured
(and not the voltage applied to the DDL). Hence, the optical
feedback removes the influence of the hysteresis. The actual
feedback signal depends on the operation mode. For the UTs, the
signal is extracted from each quadruple of ABCD images, for the ATs
from a quadruple measurement of the diodes mounted on the M3 spider
arm. In addition, there might be cases in which the science object
is bright enough to retrieve the position of the white light fringe
during the observation, e.g. when observing bright enough binaries.
Then this is the ideal feedback system. Hence, either the metrology
signal or even the actual science signal can be used to close the
loop. For the sake of loop stability we do not foresee having a
mixed sensing mode, meaning that the decision which feedback signal
to use is an either-or decision at the beginning of an observation.
The loop rate is given by the speed at which the metrology delivers
its signal, i.e. 30~Hz. This means that the loop frequency will
be in the few Hz range, i.e. the DDL loop is a relatively slow
system. This is sufficient given the maximum dOPD change rates of
31~nm/s (UTs) and 78~nm/s (ATs).
Assuming conservatively that the combination of OPD model and DDLs
will work without feedback with a precision of $10\%$ (for a short
amount of time), the maximum deviation rate that can occur is
7.8~nm/s. Then the maximum possible deviation equals the statistical
accuracy by which the phase is measured for an exposure time of
1.4~s. This will be the characteristic time scale for the control
loop. This means that the dOPD is kept at all times within $\pm
11$~nm from the nominal value (but the averaged accuracy after some
minutes is much better of course). 

The stability of the loop
crucially depends on the accuracy of the phase measurements. The
loop would break, if phase changes larger than $\pi/2$ would occur.
We require that a phase error of $\pi/4$ corresponds to a statistical fluctuation of $6\sigma$. This leads to the requirement of $\Delta \phi < \pi/4 / 6 =0.13$. As shown in section \ref{sec:phase_extraction_performance} the 
phase accuracy is always much better than that.
The loop can also monitor
the phase changes it measures, which in turn allows for a health
check.
The stability
requirement is also driving the decision whether one can track dOPD
on the science object or not. This should only be done if the
position of the white light fringe can be retrieved with sufficient
accuracy, meaning well below $\pi/2$. Another issue is mechanical
stability. The metrology system intrinsically relies on the
assumption that any mechanical disturbance does not lead to phase
changes larger than $\pi/2$. There is no way to recover this sort of
error. It can only be recovered by tracking the dOPD with the
science channel itself.

\subsection{Phase Extraction Performance}
\label{sec:phase_extraction_performance}

The measured dOPD is the product of phase and wavelength (see
section \ref{sec:phase_extraction_budget}). Therefore, the phase
recovery is an important contributor to the dOPD measurement.

\subsubsection{Theoretical Calculations}

The sensitivity of the measured phase in the ABCD algorithm to
uncorrelated noise can be calculated analytically by simple error
propagation:
\begin{equation}
\Delta \phi = \frac{1}{\sqrt{2}} \frac{\Delta I}{\langle I \rangle}
\ , \label{eq:delta_phi}
\end{equation}

where $\langle I \rangle$ is the average intensity over one ABCD
cycle and $\Delta I$ is the RMS noise of one of the intensity
measurements A, B, C and D.

The calculation above assumes completely uncorrelated errors as is
the case for the background and read-out noise of the IR camera and
IR diodes. The power output of the IR laser fluctuates with a white
noise spectrum. The temporal correlation of the fluxes results in
smaller phase errors than in the uncorrelated case. We have checked
this explicitly by simulating white noise laser light curves.
Moreover, there is a spatial correlation between the signals of the
individual pixels of the IR camera. Taking an average over the whole
pupil (the case of measuring via M2 on the UTs) yields virtually no
phase error due to intensity noise of the laser.

We conclude, that the analytical formula \ref{eq:delta_phi} is
correct for uncorrelated intensity fluctuation and is a conservative
estimate in case of a realistic noise model.

\subsubsection{UT Simulations}\label{sec:UT_simulations}

\begin{figure}[ht!]
\begin{center}
\includegraphics[totalheight=6cm]{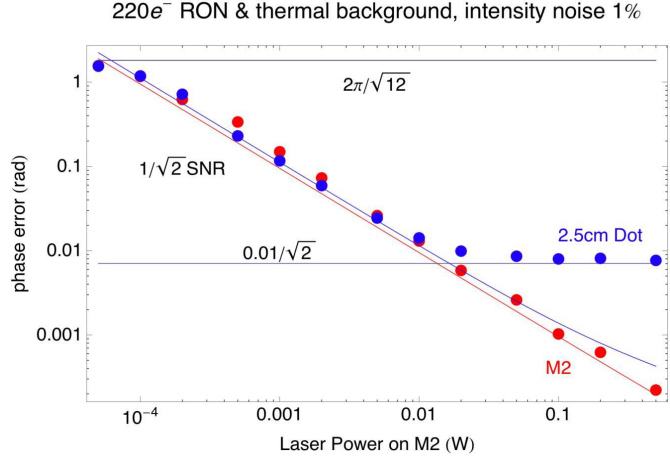}
\caption{Phase error as a function of (total) laser power on M2. The
blue dots are for the phase signal extracted from the scatterers.
The red dots use the image on M2. At low fluxes, the phase error
reaches asymptotically the value of a flat distribution. At high
fluxes, the signal from a single scatterer is dominated by the
intensity noise, which here was chosen for illustrative purpose to
be artificially high with $1\%$. The regime in which the GRAVITY
metrology will work at 4.5~mW, leading to a phase error of
0.025~rad.} \label{fig:UT_performance}
\end{center}
\end{figure}

In order to verify the metrology system design we simulated the
scattering pattern on M2 and its observation with the IR camera. In
parallel, we also simulated the signals received from the
scatterers. Thus, for each simulated image, we extracted two phase
measurements, one based on M2 and one based on the scatterers. More
specifically, the simulation parameters were:
\begin{itemize}
\item{visibility of fringe pattern: 0.9}
\item{object separation 1'', at a random angle}
\item{distance camera to M2 / scatterers: 9.5~m}
\item{IR camera: 220 electrons read-noise, $35\%$ quantum efficiency}
\item{Exposure time: 33~ms}
\item{optics: $f/0.9$, $71\%$ transmission, including laser line filter for background rejection}
\item{scatterers: $(2.5\times2.5)\ \mathrm{cm}^2$ size, albedo 0.8, scattering coefficient $2.5 \times 10^{-4} /$sr (Rabien et al.\cite{Rabien2008} 2008).}
\end{itemize}
For the phase extraction, various levels of laser power (measured as
total power arriving on M2, i.e. the sum of the two beams) have been
investigated. If no phase information could be retrieved at all due
to a too low laser power, one expects an RMS phase scatter of
$2\pi/\sqrt{12}$. At a high laser power, the intensity noise of the
laser is ultimately the limiting factor. For a single scatterer, the
phase error due to intensity noise is given by equation
\ref{eq:delta_phi}, while for the measurement via M2, the intensity
noise averages out since at each point in time many pixels sample the
full fringe pattern. In order to illustrate this, we have included
an artificially high intensity noise in the simulation. In the
regime in-between, one expects that the phase error diminishes with
$1/\sqrt{\mathrm{SNR}}$. Thus, for each power level, we generated a
set of 100 ABCD sequences and for each cycle extracted the phase in
two ways. By taking the average and RMS of each set, we were able to
measure the accuracy of the algorithm.
The final compilation of phase error versus laser power is shown in
Figure \ref{fig:UT_performance}. It fully confirms the theoretically
expected behavior at the low and high flux end. 

The chosen design yields total laser powers on M2 of 4.5mW (off-axis AO case) and 2.25mW (on-axis AO case) for UT observations, see table \ref{tab:transmission}. This corresponds to phase uncertainties per ABCD cycle of 0.025~rad and 0.05~rad. A phase error of 0.13~rad (see section \ref{sec:ddl_loop_stability}) corresponds to a total laser power of 1~mW. This is a hard limit below which the GRAVITY system would not work reliably.

\subsubsection{AT Performance}

A field of view of 8'' corresponds to 40 fringes of the metrology
laser light.  A full fringe is 50~mm wide on the 2~m diameter mirror
M1. According to table \ref{tab:transmission} the total power of the
metrology laser (two beams, red and green) at the telescope is
2.9~mW. According to section \ref{sec:calibration_diode_hardware}
we sample the fringe with an aperture
of 10~mm diameter. Therefore, we expect a flux on the diode of:
$I_A+I_B = 7.2\cdot10^{-8}$~W. The average intensity over one ABCD
phase shift cycle is then $\langle I_A+I_B \rangle = 1/2(I_A+I_B) =
3.6\cdot10^{-8}$~W . According to the specifications (see section
\ref{sec:calibration_diode_hardware}) the error of the intensity
measurement is dominated by the
noise of the amplifier of $<3\cdot10^{-12}\
\mathrm{A}/\mathrm{Hz}^{1/2}$. Together with a response of the diode
of 1 A/W this results in a noise equivalent power per 33~ms exposure
(phase shifter time step) of NEP$ = 1.7\cdot 10^{-11}$~W.

According to equation \ref{eq:delta_phi} the statistical phase error
$\Delta\phi$ per 33 ms exposure (calculated from the last 4 ABCD
measurements) is given by:
\begin{equation}
\Delta\phi = \frac{1}{\sqrt{2}} \frac{\mathrm{NEP}}{\langle I_A+I_B
\rangle} = 3.4\cdot 10^{-4} = 0.11 \ \mathrm{nm}
\frac{2\pi}{\lambda_L} \  .
\end{equation}

Therefore, the total uncertainty of the phase measurement will rather
be dominated by the uncertainties in the phase shifter calibration
and other systematic effects due to a sampling of the fringe pattern
in the pupil as compared to the case of the UTs, where the full
pupil image can be analyzed.

\subsection{Differential OPD Extraction Performance Budget}
\label{sec:phase_extraction_budget}

The performance of the instrumental differential OPD measurement is subject to the error
contributions outlined in table \ref{tab:dOPD_performance}. The metrology system fulfills the
performance requirement (section \ref{sec:accuracy_requirements}) of a phase measurement within 3 minutes to
an accuracy of 1~nm. The performance of the metrology system is
dominated by systematic effects like precision of measurement
equipment and calibration stability.

\begin{center}
\begin{table}[!ht]
\begin{center}
\small
\begin{tabular}{c c c c c}
\hline \hline
$\#$ & Phase error contribution & value &  Phase error & Phase error  \\
&&& 3 min, 2'', UT & 3 min, 2'', AT  \\
\hline
1 & wavelength uncertainty IR laser & $<30$ MHz & $<0.21$~nm & $<0.78$~nm  \\
2 & intensity noise IR laser & $<0.5\%$  per 33~ms   & $<0.04$~nm & $<0.04$~nm \\
3 & phase shifter amplitude modulation & $<0.5\%$  per 33~ms & $<0.04$~nm & $<0.04$~nm \\
4 & amplitude modulation IO coupling jitter & $<0.1\%$  per 33~ms & $<0.01$~nm & $<0.01$~nm  \\
5 & phase shifter calibration + stability & $<\lambda/5000$ &
$<0.38$~nm & $<0.38$~nm  \\
6 & amplitude variation in VLTI train & $<1\%$  per 33~ms & $<0.08$~nm
& $<0.08$~nm \\
7 & metrology zero-point calibration + stability & $<0.5$~nm &
$<0.5$~nm & $<0.5$~nm \\
8 & metrology receiver noise (UTs) & $<0.06$rad per 33~ms & $<0.5$~nm &   \\
9 & metrology receiver noise (ATs) &  $<0.01$rad per 33~ms & & $<0.11$~nm  \\
 \hline
   & Total & & $<0.84$~nm & $<1.0$~nm  \\
\hline
\end{tabular}
\caption{dOPD measurement performance error budget for the UT and AT
case.} \label{tab:dOPD_performance}
\end{center}
\end{table}
\end{center}

\section{Conclusions and Outlook}
In this document we have presented the preliminary design of a
metrology system for the GRAVITY instrument. We have shown
that the presented preliminary
design fulfills the physical requirements. With this preliminary
system design we can measure the differential phase between the
two-object beam paths with 1 nm accuracy in only 3 minutes.
Systematic errors are widely excluded, due to the mapping of the
full pupil over the full VLTI and telescope beam train. We have
defined detailed requirements for each sub-component of the
metrology system. 






\bibliography{SPIE_Metrology_2010}   
\bibliographystyle{spiebib}   

\end{document}